\documentclass[twocolumn,showpacs,preprintnumbers]{revtex4}
\usepackage{amssymb}
\usepackage{amsmath}
\usepackage{graphicx}
\usepackage{dcolumn}
\usepackage{bm}

\begin{document}

\title{Concurrence of Superpositions}
\author{Chang-shui Yu}
\author{X. X. Yi}
\author{He-shan Song}
\email{hssong@dlut.edu.cn}
\affiliation{Department of Physics, Dalian University of Technology,\\
Dalian 116024, China}
\date{\today }

\begin{abstract}
The bounds on concurrence of the superposition state in terms of those of
the states being superposed are studied in this paper. The bounds on
concurrence are quite different from those on the entanglement measure based
on von Neumann entropy (Phys. Rev. Lett. \textbf{97}, 100502 (2006)). In
particular, a \emph{nonzero} lower bound can be provided if the states being
superposed are properly constrained.
\end{abstract}

\maketitle

Most recently, Linden et al [1] have raised a problem, i.e. what is the
relation between the entanglement of two given states $\left\vert \Psi
\right\rangle $ and $\left\vert \Phi \right\rangle $ of two parties A and B
and that of their superposed state $\left\vert \Gamma \right\rangle =\alpha
\left\vert \Psi \right\rangle +\beta \left\vert \Phi \right\rangle $, with $%
\left\vert \alpha \right\vert ^{2}+\left\vert \beta \right\vert ^{2}=1$.
They have found upper bounds on the entanglement $\left\vert \Gamma
\right\rangle $ in terms of the entanglement of $\left\vert \Psi
\right\rangle $ and $\left\vert \Phi \right\rangle $, where the entanglement
measure they employed is the von Neumann entropy of the reduced state of
either of the parties [2] defined by%
\begin{equation*}
E(\Psi )=S(\text{Tr}_{A}\left\vert \Psi \right\rangle \left\langle \Psi
\right\vert )=S(\text{Tr}_{B}\left\vert \Psi \right\rangle \left\langle \Psi
\right\vert ).
\end{equation*}

Since the entanglement measure for pure states is not unique, it is natural
to ask whether the bounds obtained in Ref. [1] only exist for von Neumann
entropy? Motivated by this question, in the paper we employ concurrence
[3,4,5] as the entanglement measure to study how the concurrence of $%
\left\vert \Gamma \right\rangle $ is bounded by the concurrence of $%
\left\vert \Psi \right\rangle $ and $\left\vert \Phi \right\rangle $. The
result shows that bounds on entanglement of superposition depend on the
entanglement measure. Even though the form of the bounds for concurrence are
something like those given in Ref. [1], they are quite different. For
example, for two biorthogonal states, Ref. [1] has shown an elegant bound
for the von Neumann entropy of their superposition, i.e. an equality bound,
while there do not exist explicit constraints for the two states such that
the concurrence of their superposition has equality bounds. It is the most
important that a \textit{nonzero} lower bound on the concurrence of a
superposition state can be provided if the states being superposed satisfy
some conditions which includes the constraints on the concurrence of the
states and the their proportions in the superposition state and so on. The
paper is organized as follows. First, we introduce an variational but
equivalent expression for concurrence; Then we study the concurrence of
superposition by the analogous logic to that of Ref. [1]; The conclusion is
drawn at last.

In this paragraph, we first introduce the concurrence and derive the
variational form of concurrence which will simplify our presentation. As we
know, $\left\vert \psi \right\rangle _{AB}$ of two parties A and B defined
in $\left( n_{1}\times n_{2}\right) $ dimension can, in general, be
considered as a vector, i.e. $\left\vert \psi \right\rangle
_{AB}=[a_{00},a_{01},\cdot \cdot \cdot ,a_{0n_{2}},a_{10},a_{11},\cdot \cdot
\cdot ,a_{n_{1}n_{2}}]^{T}$ with superscript $T$ denoting transpose
operation, while throughout of the paper we confine all the pure states to
matrix notation, i.e.
\begin{equation}
\psi =\left(
\begin{array}{cccc}
a_{00} & a_{01} & \cdots & a_{0n_{2}} \\
a_{10} & a_{11} & \cdots & a_{1n_{2}} \\
\vdots & \vdots & \ddots & \vdots \\
a_{n_{1}0} & a_{n_{1}1} & \cdots & a_{n_{1}n_{2}}%
\end{array}%
\right) .
\end{equation}%
With the matrix notation, one can easily find that the reduced density
matrix
\begin{equation}
\rho _{B}=\psi \psi ^{\dag }.
\end{equation}%
Consider the eigenvalue decomposition of $\rho _{B}$, one can have
\begin{equation}
\rho _{B}=\psi \psi ^{\dag }=\Psi M\Psi ^{\dag },
\end{equation}%
where the columns of $\Psi $ correspond to the eigenvectors of $\rho _{B}$
and $M$ is a non-negative diagonal matrix with its diagonal entries
corresponding to the eigenvalues of $\rho _{B}$ or the square of the
singular values of $\psi $.

The concurrence for an arbitrary dimensional bipartite pure state $%
\left\vert \psi \right\rangle $ is defined [4] by
\begin{equation}
C(\left\vert \psi \right\rangle )=\sqrt{1-\text{Tr}(\rho _{r}^{2})},
\end{equation}%
which turned out to be the length of the concurrence vector by Wootters [6],
where $\rho _{r}=Tr_{\alpha }\left\vert \psi \right\rangle \left\langle \psi
\right\vert $ denoting the reduced density matrix tracing over either of the
two parties. Substitute eq. (3) into eq. (4), we have
\begin{eqnarray}
C(\left\vert \psi \right\rangle ) &=&\sqrt{1-\sum\limits_{i}\sigma _{i}^{4}}
\\
&=&\sqrt{\sum\limits_{i\neq j}\sigma _{i}^{2}\sigma _{j}^{2}},
\end{eqnarray}%
where $\sigma _{i}$, $\sum_{i}\sigma _{i}^{2}=1$, is one singular value of $%
\psi $ where the normalized $\psi $ is always implied. Eq. (5) and eq. (6)
are the so-called variational forms for concurrence to be used in the paper.

\textbf{Theorem 1: }If the two pure states $\Psi _{1}$ and $\Phi _{1}$
defined in $\left( n\times m\right) $ dimension satisfy $\Psi _{1}\Phi
_{1}^{\dagger }=0$, the concurrence of their superposed states $\Gamma
_{1}^{+}=\alpha \Psi _{1}+\beta \Phi _{1}$ with $\left\vert \alpha
\right\vert ^{2}+\left\vert \beta \right\vert ^{2}=1,$ obeys%
\begin{equation*}
\frac{\left\vert \alpha \right\vert ^{2}C(\Psi _{1})+\left\vert \beta
\right\vert ^{2}C(\Phi _{1})}{2}\leq C(\alpha \Psi _{1}+\beta \Phi _{1})
\end{equation*}%
\begin{equation}
\leq \frac{\left\vert \alpha \right\vert ^{2}\tilde{C}(\Psi _{1},\alpha
)+\left\vert \beta \right\vert ^{2}\tilde{C}(\Phi _{1},\beta )}{2},
\end{equation}%
where%
\begin{equation}
\tilde{C}(\Psi _{1},\alpha )=\sqrt{C^{2}(\Psi _{1})+\frac{\left\vert \beta
\right\vert ^{4}}{\left\vert \alpha \right\vert ^{4}}+2\frac{\left\vert
\beta \right\vert ^{2}}{\left\vert \alpha \right\vert ^{2}}},
\end{equation}%
with $\left\vert \beta \right\vert ^{2}=1-\left\vert \alpha \right\vert ^{2}$%
.

That we say $\Psi _{1}$ and $\Phi _{1}$ defined in the same dimension
implies that the two states have been properly adjusted. Note that $\Psi
_{1} $ and $\Phi _{1}$ may be defined in the Hilbert space with different
dimensions. However, one can always add some zero entries to $\Psi _{1}$ and
$\Phi _{1}$ such that $\Psi _{1}$ and $\Phi _{1}$ are defined in the same
dimension.

\textbf{Proof:} As we know, for any two Hermitian matrix $H$ and $K$ defined
in $C^{n\times n}$,
\begin{equation}
\lambda _{i}(H)+\lambda _{1}(K)\leq \lambda _{i}(H+K)\leq \lambda
_{i}(H)+\lambda _{n}(K)
\end{equation}%
holds, where $\lambda _{i}(\cdot )$ denotes the eigenvalues in increasing
order [7] (See \textbf{Theorem 4.3.1} in Ref. [7]).

Since $\Psi _{1}\Psi _{1}^{\dagger }$ and $\Phi _{1}\Phi _{1}^{\dagger }$
are both Hermitian and defined in $\left( n\times n\right) $ dimension, one
has
\begin{eqnarray}
&&\left\vert \alpha \right\vert ^{2}\lambda _{i}(\Psi _{1}\Psi _{1}^{\dagger
})+\left\vert \beta \right\vert ^{2}\lambda _{1}(\Phi _{1}\Phi _{1}^{\dagger
})  \notag \\
&\leq &\lambda _{i}(\left\vert \alpha \right\vert ^{2}\Psi _{1}\Psi
_{1}^{\dagger }+\left\vert \beta \right\vert ^{2}\Phi _{1}\Phi _{1}^{\dagger
}).
\end{eqnarray}%
Because $\Psi _{1}\Phi _{1}^{\dagger }=0$,
\begin{equation}
\lambda _{i}(\left\vert \alpha \right\vert ^{2}\Psi _{1}\Psi _{1}^{\dagger
}+\left\vert \beta \right\vert ^{2}\Phi _{1}\Phi _{1}^{\dagger })=\lambda
_{i}\left( \Gamma _{1}^{+}\left( \Gamma _{1}^{+}\right) ^{\dag }\right) .
\end{equation}%
Substitute eq. (11) into eq. (6), we have
\begin{eqnarray}
&&\left\{ \sum\limits_{i\neq j}^{n}\left[ \left\vert \alpha \right\vert
^{2}\lambda _{i}(\Psi _{1}\Psi _{1}^{\dagger })+\left\vert \beta \right\vert
^{2}\lambda _{1}(\Phi _{1}\Phi _{1}^{\dagger })\right] \right.  \notag \\
&&\times \left. \left[ \left\vert \alpha \right\vert ^{2}\lambda _{j}(\Psi
_{1}\Psi _{1}^{\dagger })+\left\vert \beta \right\vert ^{2}\lambda _{1}(\Phi
_{1}\Phi _{1}^{\dagger })\right] \right\} ^{1/2}  \notag \\
&=&\left\{ \left\vert \alpha \right\vert ^{4}C^{2}(\Psi
_{1})+(n-1)\left\vert \beta \right\vert ^{2}\lambda _{1}(\Phi _{1}\Phi
_{1}^{\dagger })\right.  \notag \\
&&\times \left. \left[ 2\left\vert \alpha \right\vert ^{2}+n\left\vert \beta
\right\vert ^{2}\lambda _{1}(\Phi _{1}\Phi _{1}^{\dagger })\right] \right\}
^{\frac{1}{2}}  \notag \\
&\leq &\sqrt{\sum\limits_{i\neq j}^{n}\lambda _{i}\left( \Gamma
_{1}^{+}\left( \Gamma _{1}^{+}\right) \right) \lambda _{j}\left( \Gamma
_{1}^{+}\left( \Gamma _{1}^{+}\right) \right) }=C(\Gamma _{1}^{+}).
\end{eqnarray}%
Substitute eq. (11) into eq. (5), we have
\begin{eqnarray}
C(\Gamma _{1}^{+}) &=&\sqrt{1-\sum\limits_{i}^{n}\lambda _{i}^{2}\left(
\Gamma _{1}^{+}\left( \Gamma _{1}^{+}\right) \right) }  \notag \\
&\leq &\left[ \left\vert \alpha \right\vert
^{4}(1-\sum\limits_{i}^{n}\lambda _{i}^{2}(\Psi _{1}\Psi _{1}^{\dagger
}))\right. +\left\vert \beta \right\vert ^{4}(1-n\lambda _{1}^{2}(\Phi
_{1}\Phi _{1}^{\dagger }))  \notag \\
&&+\left. 2\left\vert \alpha \right\vert ^{2}\left\vert \beta \right\vert
^{2}(1-\lambda _{1}\sum\limits_{i}^{n}\lambda _{i}(\Psi _{1}\Psi
_{1}^{\dagger }))\right] ^{\frac{1}{2}}  \notag \\
&=&\left[ \left\vert \alpha \right\vert ^{4}C^{2}(\Psi _{1})+\left\vert
\beta \right\vert ^{4}(1-n\lambda _{1}^{2}(\Phi _{1}\Phi _{1}^{\dagger
}))\right.  \notag \\
&&\left. +2\left\vert \alpha \right\vert ^{2}\left\vert \beta \right\vert
^{2}(1-\lambda _{1}(\Phi _{1}\Phi _{1}^{\dagger }))\right] ^{\frac{1}{2}}.
\end{eqnarray}%
Simplifying eq. (12) and eq. (13) by considering the positive semidefinite $%
\Phi \Phi ^{\dagger }$, the two equations can be rewritten by

\begin{equation}
\left\vert \alpha \right\vert ^{2}C(\Psi _{1})\leq C(\alpha \Psi _{1}+\beta
\Phi _{1})\leq \sqrt{\left\vert \alpha \right\vert ^{4}C^{2}(\Psi
_{1})+\left\vert \beta \right\vert ^{4}+2\left\vert \alpha \right\vert
^{2}\left\vert \beta \right\vert ^{2}}.
\end{equation}%
Consider the analogous relation to eq. (9) by exchanging $H$ and $K$ and the
positive semidefinite $\Psi \Psi ^{\dagger }$, based on the above procedure
one can also obtain
\begin{equation}
\left\vert \beta \right\vert ^{2}C(\Phi _{1})\leq C(\alpha \Psi _{1}+\beta
\Phi _{1})\leq \sqrt{\left\vert \beta \right\vert ^{4}C^{2}(\Phi
_{1})+\left\vert \alpha \right\vert ^{4}+2\left\vert \alpha \right\vert
^{2}\left\vert \beta \right\vert ^{2}}.
\end{equation}%
Therefore, eq. (14) and eq. (15) can be given in a symmetric form by%
\begin{equation*}
\frac{\left\vert \alpha \right\vert ^{2}C(\Psi _{1})+\left\vert \beta
\right\vert ^{2}C(\Phi _{1})}{2}\leq C(\alpha \Psi _{1}+\beta \Phi _{1})
\end{equation*}%
\begin{equation}
\leq \frac{\left\vert \alpha \right\vert ^{2}\tilde{C}(\Psi _{1},\alpha
)+\left\vert \beta \right\vert ^{2}\tilde{C}(\Phi _{1},\beta )}{2},
\end{equation}%
which completes the proof.$\Box $

\textbf{Theorem 2: }If $\left( n\times m\right) $-dimensional pure states $%
\Phi _{2}$ and $\Psi _{2}$ are orthogonal, i.e. Tr$\Psi _{2}\Phi
_{2}^{\dagger }=0$, the concurrence of their superposed state $\Gamma
_{2}^{+}=\alpha \Psi _{2}+\beta \Phi _{2}$ with rank $r$ satisfy:%
\begin{eqnarray}
&&\left[ \left\vert \alpha \right\vert ^{2}l\left( \alpha ,\Psi _{2},\Phi
_{2}\right) +\left\vert \beta \right\vert ^{2}l\left( \beta ,\Phi _{2},\Psi
_{2}\right) \right]  \notag \\
&\leq &2\max \{\left\vert \alpha \right\vert ^{2}l\left( \alpha ,\Psi
_{2},\Phi _{2}\right) ,\left\vert \beta \right\vert ^{2}l\left( \beta ,\Phi
_{2},\Psi _{2}\right) \}  \notag \\
&\leq &C(\Gamma _{2}^{+})\leq 2\min \{\left\vert \alpha \right\vert
^{2}f\left( \alpha ,\Psi _{2},\Phi _{2}\right) ,\left\vert \beta \right\vert
^{2}f\left( \beta ,\Phi _{2},\Psi _{2}\right) \}  \notag \\
&\leq &\left[ \left\vert \alpha \right\vert ^{2}f\left( \alpha ,\Psi
_{2},\Phi _{2}\right) +\left\vert \beta \right\vert ^{2}f\left( \beta ,\Phi
_{2},\Psi _{2}\right) \right] ,
\end{eqnarray}%
where
\begin{eqnarray*}
&&l\left( \alpha ,\Psi _{2},\Phi _{2}\right) \\
&=&\left[ \max \left\{ 0,\right. C^{2}\left( \Psi _{2}\right) +\frac{%
\left\vert \beta \right\vert ^{4}}{\left\vert \alpha \right\vert ^{4}}\left[
1-r\lambda _{n}^{2}(\Phi _{2}\Phi _{2}^{\dagger })\right] \right. \\
&&\left. \left. +2\frac{\left\vert \beta \right\vert ^{2}}{\left\vert \alpha
\right\vert ^{2}}\lambda _{n}(\Phi _{2}\Phi _{2}^{\dagger })-\frac{3}{%
4\left\vert \alpha \right\vert ^{4}}\right\} \right] ^{\frac{1}{2}},
\end{eqnarray*}%
and%
\begin{eqnarray*}
f\left( \alpha ,\Psi _{2},\Phi _{2}\right) &=&\left\{ C^{2}\left( \Psi
_{2}\right) +(r-1)\frac{\left\vert \beta \right\vert ^{2}}{\left\vert \alpha
\right\vert ^{2}}\lambda _{n}(\Phi _{2}\Phi _{2}^{\dagger })\right. \\
&&\times \left. \left[ 2+r\frac{\left\vert \beta \right\vert ^{2}}{%
\left\vert \alpha \right\vert ^{2}}\lambda _{n}(\Phi _{2}\Phi _{2}^{\dagger
})\right] \right\} ^{\frac{1}{2}}.
\end{eqnarray*}%
\textbf{Proof.} Construct matrix $D_{2}$ such that
\begin{equation*}
D_{2}=\Phi _{2}\Phi _{2}^{\dagger }+\Psi _{2}\Psi _{2}^{\dagger }.
\end{equation*}%
The inequality (10) holds in this case, too. I.e.
\begin{eqnarray}
&&\left\vert \alpha \right\vert ^{2}\lambda _{i}(\Psi _{2}\Psi _{2}^{\dagger
})+\left\vert \beta \right\vert ^{2}\lambda _{2}(\Phi _{2}\Phi _{2}^{\dagger
})  \notag \\
&\leq &\lambda _{i}(D_{2})\leq \left\vert \alpha \right\vert ^{2}\lambda
_{i}(\Psi _{2}\Psi _{2}^{\dagger })+\left\vert \beta \right\vert ^{2}\lambda
_{n}(\Phi _{2}\Phi _{2}^{\dagger })
\end{eqnarray}%
$D_{2}$ can also be rewritten as%
\begin{equation}
D_{2}=\frac{1}{2}\Gamma _{2}^{+}\left( \Gamma _{2}^{+}\right) ^{\dag }+\frac{%
1}{2}\Gamma _{2}^{-}\left( \Gamma _{2}^{-}\right) ^{\dag },
\end{equation}%
with $\Gamma _{2}^{-}=\alpha \Psi _{2}-\beta \Phi _{2}$. In terms of
inequality (9), eq. (19) follows that
\begin{eqnarray}
&&\frac{1}{2}\lambda _{i}\left( (\Gamma _{2}^{+}\left( \Gamma
_{2}^{+}\right) ^{\dag }\right) +\frac{1}{2}\lambda _{1}(\Gamma
_{2}^{-}\left( \Gamma _{2}^{-}\right) ^{\dag })  \notag \\
&\leq &\lambda _{i}(D_{2})\leq \frac{1}{2}\lambda _{i}(\Gamma _{2}^{+}\left(
\Gamma _{2}^{+}\right) ^{\dag })+\frac{1}{2}\lambda _{n}(\Gamma
_{2}^{-}\left( \Gamma _{2}^{-}\right) ^{\dag }).
\end{eqnarray}%
Comparing eq. (18) with eq. (20), one has%
\begin{eqnarray}
&&\frac{1}{2}\lambda _{i}(\Gamma _{2}^{+}\left( \Gamma _{2}^{+}\right)
^{\dag })+\frac{1}{2}\lambda _{1}(\Gamma _{2}^{-}\left( \Gamma
_{2}^{-}\right) ^{\dag })  \notag \\
&\leq &\left\vert \alpha \right\vert ^{2}\lambda _{i}(\Psi _{2}\Psi
_{2}^{\dagger })+\left\vert \beta \right\vert ^{2}\lambda _{n}(\Phi _{2}\Phi
_{2}^{\dagger }),
\end{eqnarray}%
and
\begin{eqnarray}
&&\frac{1}{2}\lambda _{i}(\Gamma _{2}^{+}\left( \Gamma _{2}^{+}\right)
^{\dag })+\frac{1}{2}\lambda _{1}(\Gamma _{2}^{-}\left( \Gamma
_{2}^{-}\right) ^{\dag })  \notag \\
&\leq &\left\vert \alpha \right\vert ^{2}\lambda _{n}(\Psi _{2}\Psi
_{2}^{\dagger })+\left\vert \beta \right\vert ^{2}\lambda _{i}(\Phi _{2}\Phi
_{2}^{\dagger }).
\end{eqnarray}%
Due to the positive semidefinite $\Gamma _{2}^{-}\left( \Gamma
_{2}^{-}\right) ^{\dag }$, $\lambda _{1}(\Gamma _{2}^{-}\left( \Gamma
_{2}^{-}\right) ^{\dag })\geqslant 0$. Eq. (21) and eq. (22) can be
rewritten by
\begin{equation}
\frac{1}{2}\lambda _{i}(\Gamma _{2}^{+}\left( \Gamma _{2}^{+}\right) ^{\dag
})\leq \left\vert \alpha \right\vert ^{2}\lambda _{i}(\Psi _{2}\Psi
_{2}^{\dagger })+\left\vert \beta \right\vert ^{2}\lambda _{n}(\Phi _{2}\Phi
_{2}^{\dagger })
\end{equation}%
and
\begin{equation}
\frac{1}{2}\lambda _{i}(\Gamma _{2}^{+}\left( \Gamma _{2}^{+}\right) ^{\dag
})\leq \left\vert \alpha \right\vert ^{2}\lambda _{n}(\Psi _{2}\Psi
_{2}^{\dagger })+\left\vert \beta \right\vert ^{2}\lambda _{i}(\Phi _{2}\Phi
_{2}^{\dagger }).
\end{equation}%
Substitute eq. (23) and eq. (24) into eq. (6), we arrive at

\begin{equation}
\frac{1}{2}C(\Gamma _{2}^{+})\leq \left\vert \alpha \right\vert ^{2}f\left(
\alpha ,\Psi _{2},\Phi _{2}\right)
\end{equation}%
and
\begin{equation}
\frac{1}{2}C(\Gamma _{2}^{+})\leq \left\vert \beta \right\vert ^{2}f\left(
\beta ,\Phi _{2},\Psi _{2}\right) .
\end{equation}%
Rewriting eq. (25) and eq. (26) in a symmetric form, it follows that%
\begin{eqnarray}
C(\Gamma _{2}^{+}) &\leq &2\min \{\left\vert \alpha \right\vert ^{2}f\left(
\alpha ,\Psi _{2},\Phi _{2}\right) ,\left\vert \beta \right\vert ^{2}f\left(
\beta ,\Phi _{2},\Psi _{2}\right) \}  \notag \\
&\leq &\left[ \left\vert \alpha \right\vert ^{2}f\left( \alpha ,\Psi
_{2},\Phi _{2}\right) +\left\vert \beta \right\vert ^{2}f\left( \beta ,\Phi
_{2},\Psi _{2}\right) \right] .
\end{eqnarray}%
According to eq. (5) and eq. (23), we have
\begin{equation*}
\frac{1}{2}C(\Gamma _{2}^{+})=\sqrt{\frac{1}{4}-\frac{1}{4}\sum_{i}\lambda
_{i}^{2}(\Gamma _{2}^{+}\left( \Gamma _{2}^{+}\right) ^{\dag })}
\end{equation*}%
\begin{eqnarray}
&\geq &\sqrt{\max \left\{ 0,\frac{1}{4}-\sum_{i}\left[ \left\vert \alpha
\right\vert ^{2}\lambda _{i}(\Psi _{2}\Psi _{2}^{\dagger })+\left\vert \beta
\right\vert ^{2}\lambda _{n}(\Phi _{2}\Phi _{2}^{\dagger })\right]
^{2}\right\} }  \notag \\
&=&\left\vert \alpha \right\vert ^{2}l\left( \alpha ,\Psi _{2},\Phi
_{2}\right) .
\end{eqnarray}%
It is obvious that eq. (28) can not always provide a good (\textit{nonzero})
lower bound for all the cases. One can find that if and only if
\begin{equation}
\left\vert \beta \right\vert ^{4}+\left\vert \alpha \right\vert ^{4}\left[
C^{2}\left( \Psi _{2}\right) +\frac{1}{r}\right] >\frac{3}{4},
\end{equation}%
there may exist some $\Phi _{2}$ such that eq. (28) can provide a good lower
bound. Analogously, we can also obtain
\begin{equation}
\frac{1}{2}C(\Gamma _{2}^{+})\geq \left\vert \beta \right\vert ^{2}l\left(
\beta ,\Phi _{2},\Psi _{2}\right) ,
\end{equation}%
where a similar condition of $\Phi _{2}$ to eq. (29) is needed for $\Psi
_{2} $ to give a good lower bound. Hence, one can obtain a symmetric form
\begin{eqnarray}
\frac{1}{2}C(\Gamma _{2}^{+}) &\geq &\max \left\{ \left\vert \alpha
\right\vert ^{2}l\left( \alpha ,\Psi _{2},\Phi _{2}\right) ,\left\vert \beta
\right\vert ^{2}l\left( \beta ,\Phi _{2},\Psi _{2}\right) \right\}  \notag \\
&\geq &\frac{1}{2}\left[ \left\vert \alpha \right\vert ^{2}l\left( \alpha
,\Psi _{2},\Phi _{2}\right) +\left\vert \beta \right\vert ^{2}l\left( \beta
,\Phi _{2},\Psi _{2}\right) \right] .
\end{eqnarray}%
A good bound requires that the condition (29) holds at least for one of the
states being superposed. $\Box $

\textbf{Theorem 3}: For any two normalized $\left( n\times \ m\right) $%
-dimensional pure states $\Psi _{3}$ and $\Phi _{3}$ with $\left\vert \alpha
\right\vert ^{2}+\left\vert \beta \right\vert ^{2}=1$, the concurrence of
the superposed state $\Gamma _{3}^{+}=\alpha \Psi _{3}+\beta \Phi _{3}$ with
rank $r$ is bounded as%
\begin{eqnarray*}
&&\frac{1}{2}\left[ \left\vert \alpha \right\vert ^{2}\tilde{l}\left( \alpha
,\Psi _{3},\Phi _{3}\right) +\left\vert \beta \right\vert ^{2}\tilde{l}%
\left( \beta ,\Phi _{3},\Psi _{3}\right) \right] \\
&\leq &\max \left\{ \left\vert \alpha \right\vert ^{2}\tilde{l}\left( \alpha
,\Psi _{3},\Phi _{3}\right) ,\left\vert \beta \right\vert ^{2}\tilde{l}%
\left( \beta ,\Phi _{3},\Psi _{3}\right) \right\} \\
&\leq &\frac{\left\Vert \Gamma _{3}^{+}\right\Vert ^{2}}{2}C(\Gamma
_{3}^{+})\leq \min \{\left\vert \alpha \right\vert ^{2}f\left( \alpha ,\Psi
_{3},\Phi _{3}\right) ,\left\vert \beta \right\vert ^{2}f\left( \beta ,\Phi
_{3},\Psi _{3}\right) \}
\end{eqnarray*}%
\begin{equation}
\leq \frac{1}{2}\left[ \left\vert \alpha \right\vert ^{2}f\left( \alpha
,\Psi _{3},\Phi _{3}\right) +\left\vert \beta \right\vert ^{2}f\left( \beta
,\Phi _{3},\Psi _{3}\right) \right] ,
\end{equation}%
where
\begin{equation*}
\tilde{l}\left( \alpha ,\Psi _{3},\Phi _{3}\right)
\end{equation*}%
\begin{eqnarray*}
&=&\left[ \max \left\{ 0,\right. C^{2}\left( \Psi _{3}\right) +\frac{%
\left\vert \beta \right\vert ^{4}}{\left\vert \alpha \right\vert ^{4}}\left[
1-r\lambda _{n}^{2}(\Phi _{3}\Phi _{3}^{\dagger })\right] \right. \\
&&\left. \left. +2\frac{\left\vert \beta \right\vert ^{2}}{\left\vert \alpha
\right\vert ^{2}}\lambda _{n}(\Phi _{3}\Phi _{3}^{\dagger })-\frac{1}{%
\left\vert \alpha \right\vert ^{4}}\left( 1-\frac{\left\Vert \Gamma
_{3}^{+}\right\Vert ^{4}}{4}\right) \right\} \right] ^{\frac{1}{2}},
\end{eqnarray*}

\textbf{Proof.} Analogous to theorem 2, consider the matrix
\begin{equation}
D_{3}=\left\vert \alpha \right\vert ^{2}\Psi _{3}\Psi _{3}^{\dagger
}+\left\vert \beta \right\vert ^{2}\Phi _{3}\Phi _{3}^{\dagger }.
\end{equation}%
$D_{3}$ can be rewritten as
\begin{equation}
D_{3}=\frac{\left\Vert \Gamma _{3}^{+}\right\Vert ^{2}}{2}\tilde{\Gamma}%
_{3}^{+}\left( \tilde{\Gamma}_{3}^{+}\right) ^{\dagger }+\frac{\left\Vert
\Gamma _{3}^{-}\right\Vert ^{2}}{2}\tilde{\Gamma}_{3}^{-}\left( \tilde{\Gamma%
}_{3}^{-}\right) ^{\dagger },
\end{equation}%
with $\Gamma _{3}^{\pm }=\alpha \Psi _{3}\pm \beta \Phi _{3}$ and $\tilde{%
\Gamma}_{3}^{\pm }=\frac{\Gamma _{3}^{\pm }}{\left\Vert \Gamma _{3}^{\pm
}\right\Vert }$. Based on eq. (9), we have%
\begin{eqnarray}
&&\frac{\left\Vert \Gamma _{3}^{+}\right\Vert ^{2}}{2}\lambda _{i}\left( (%
\tilde{\Gamma}_{3}^{+}\left( \tilde{\Gamma}_{3}^{+}\right) ^{\dagger
}\right) +\frac{\left\Vert \Gamma _{3}^{-}\right\Vert ^{2}}{2}\lambda
_{1}\left( \tilde{\Gamma}_{3}^{-}\left( \tilde{\Gamma}_{3}^{-}\right)
^{\dagger }\right)  \notag \\
&\leq &\lambda _{i}(D_{3})\leq \left\vert \alpha \right\vert ^{2}\lambda
_{i}\left( \Psi _{3}\Psi _{3}^{\dagger }\right) +\left\vert \beta
\right\vert ^{2}\lambda _{1}\left( \Phi _{3}\Phi _{3}^{\dagger }\right) .
\end{eqnarray}%
Following the similar procedure to that of theorem 2, based on eq. (35) one
can obtain
\begin{equation*}
\frac{\left\Vert \Gamma _{3}^{+}\right\Vert ^{2}}{2}C\left( \Gamma
_{3}^{+}\right) \leq \min \{\left\vert \alpha \right\vert ^{2}f\left( \alpha
,\Psi _{3},\Phi _{3}\right) ,\left\vert \beta \right\vert ^{2}f\left( \beta
,\Phi _{3},\Psi _{3}\right) \}
\end{equation*}%
\begin{equation}
\leq \frac{1}{2}\left[ \left\vert \alpha \right\vert ^{2}f\left( \alpha
,\Psi _{3},\Phi _{3}\right) +\left\vert \beta \right\vert ^{2}f\left( \beta
,\Phi _{3},\Psi _{3}\right) \right] .
\end{equation}%
Note that $C(\Gamma _{3}^{+})$ means the concurrence of the normalized $%
\alpha \Psi _{3}+\beta \Phi _{3}$, i.e. $C(\tilde{\Gamma}_{3}^{+})$. From
eq. (5) and eq. (35) again, one has

\begin{equation*}
\frac{\left\Vert \Gamma _{3}^{+}\right\Vert ^{2}}{2}C\left( \Gamma
_{3}^{+}\right) =\sqrt{\frac{\left\Vert \Gamma _{3}^{+}\right\Vert ^{4}}{4}-%
\frac{\left\Vert \Gamma _{3}^{+}\right\Vert ^{4}}{4}\lambda _{i}^{2}\left( (%
\tilde{\Gamma}_{3}^{+}\left( \tilde{\Gamma}_{3}^{+}\right) ^{\dagger
}\right) }
\end{equation*}%
\begin{equation*}
\geq \sqrt{\max \left\{ 0,\frac{\left\Vert \Gamma _{3}^{+}\right\Vert ^{4}}{4%
}-\sum_{i}\left[ \left\vert \alpha \right\vert ^{2}\lambda _{i}(\Psi
_{3}\Psi _{3}^{\dagger })+\left\vert \beta \right\vert ^{2}\lambda _{n}(\Phi
_{3}\Phi _{3}^{\dagger })\right] ^{2}\right\} }
\end{equation*}%
\begin{equation}
=\left\vert \alpha \right\vert ^{2}\tilde{l}(\alpha ,\Psi _{3},\Phi _{3}).
\end{equation}%
Analogously, if and only if
\begin{equation}
\left\vert \beta \right\vert ^{4}+\left\vert \alpha \right\vert ^{4}\left[
C^{2}\left( \Psi _{3}\right) +\frac{1}{r}\right] >1-\frac{\left\Vert \Gamma
_{3}^{+}\right\Vert ^{4}}{4},
\end{equation}%
there may exist some $\Phi _{3}$ such that eq. (37) can give a nonzero lower
bound. In a symmetric form, the bound on concurrence can be given by%
\begin{eqnarray}
&&\frac{1}{2}\left[ \left\vert \alpha \right\vert ^{2}\tilde{l}\left( \alpha
,\Psi _{3},\Phi _{3}\right) +\left\vert \beta \right\vert ^{2}\tilde{l}%
\left( \beta ,\Phi _{3},\Psi _{3}\right) \right]   \notag \\
&\leq &\max \left\{ \left\vert \alpha \right\vert ^{2}\tilde{l}\left( \alpha
,\Psi _{3},\Phi _{3}\right) ,\left\vert \beta \right\vert ^{2}\tilde{l}%
\left( \beta ,\Phi _{3},\Psi _{3}\right) \right\}   \notag \\
&\leq &\frac{\left\Vert \Gamma _{3}^{+}\right\Vert ^{2}}{2}C\left( \Gamma
_{3}^{+}\right) .
\end{eqnarray}%
In order to obtain a good lower bound, the analogous conditions to eq. (29)
is needed at least for one of $\Psi _{3}\ $and $\Phi _{3}$. Eq. (36) and eq.
(37) complete the proof.$\Box $

In summary, we have given the bounds on the concurrence of superposition
states, which are very different from those in Ref. [1]. A lower bound can
also be provided if the states being superposed are constrained as
mentioned. However, it seems to be difficult to present a useful lower bound
for arbitrary two states by the current approach. As to the superposition of
more than two terms, one has to repeat our procedure based on eq. (9). One
can easily see that if the current bound on concurrence is converted into
that on the square of concurrence (it is only a simple algebra), the
generalization to the case of more than two terms will be straight and
convenient. What is more, one will see that if the negativity [8] is
employed as entanglement measure, it is also difficult to find useful (upper
and lower) bounds based on the current approach.

This work was supported by the National Natural Science Foundation of China,
under Grant Nos. 10575017 and 60472017.


\begin{thebibliography}{9}
\bibitem{[1]} Noah Linden, Sandu Popescu and John A. Smolin, Phys. Rev.
Lett. \textbf{97}, 100502 (2006).

\bibitem{[2]} C. H. Bennett, H. J. Bernstein, S. Popescu, and B. Schumacher,
Phys. Rev. A \textbf{53}, 2046 (1996).

\bibitem{[3]} W. K. Wootters, Phys. Rev. Lett. \textbf{80}, 2245 (1998).

\bibitem{[4]} Pranaw Rungta, V. Bu\v{z}ek, Carlton M. Caves, M. Hillery, and
G. J. Milburn, Phy.Rev. A\textbf{\ 64}, 042315 (2001).

\bibitem{[5]} K. Audenaert, F.Verstraete and De Moor, Phys. Rev. A \textbf{64%
}, 052304 (2001)

\bibitem{[6]} W. K. Wootters, Quantum Inf. Comp. \textbf{1}, 27 (2001).

\bibitem{[7]} R. A. Horn and C. R. Johnson, \textit{Matrix Analysis }%
(Cambridge University Press, New York, 1985).

\bibitem{[8]} G. Vidal and R. F. Werner, Phys. Rev. A \textbf{65}, 032314
(2002).
\end{thebibliography}
\end{document}